# Superconducting transition in Nb nanowires fabricated using focused ion beam

G C Tettamanzi[1,2]*, C I Pakes[3], A Potenza[4], S. Rubanov[5], C H Marrows[4] and S Prawer[1]

[1] School of Physics, University of Melbourne, Victoria 3010, Australia

[2] Kavli Institute of Nanoscience, TU Delft, Lorentzweg 1, 2628 CJ DELFT, The Netherlands

[3] Department of Physics, La Trobe University, Victoria 3086, Australia

[4] School of Physics and Astronomy, University of Leeds, Leeds LS2 9JT, United Kingdom

[5] Bio21 Institute, University of Melbourne, Australia

*E-mail: G.C.Tettamanzi@tudelft.nl

**Abstract.** Making use of focused Ga-ion beam (FIB) fabrication technology, the evolution with device dimension of the low-temperature electrical properties of Nb nanowires has been examined in a regime where crossover from Josephson-like to insulating behaviour is evident. Resistance-temperature data for devices with a physical width of order 100 nm demonstrate suppression of superconductivity, leading to dissipative behaviour that is shown to be consistent with the activation of phase-slip below $T_c$. This study suggests that by exploiting the Ga-impurity poisoning introduced by the FIB into the periphery of the nanowire, a central superconducting phase-slip nanowire with sub-10 nm dimensions may be engineered within the core of the nanowire.

## 1. Introduction

The use of focused ion beam technology to fabricate nanoscale superconducting components has developed recent interest, particularly as an approach to engineer miniature, low-noise SQUID technology based upon nanowire junctions [1-3]. However, the manner in which superconductivity is suppressed as the nanowire width is reduced has not been explored for low-$T_c$ FIB-engineered devices. This is of importance in assessing the extent to which FIB-patterned components can be further miniaturized for superconducting nano-electronics and motivated particularly by the possibility of introducing new device functionality that exploits phase-slip processes driven by macroscopic quantum tunneling in the one-dimensional limit [4].

For superconducting components with dimensions on the order of the Ginzburg-Landau (GL) coherence length, local fluctuations may lead to dissipation at temperatures below the superconducting transition temperature, $T_c$ [5]. A residual resistance close to $T_c$ has been extensively reported [4], and is understood within a theoretical framework of phase-slip arising from thermal activation over the free energy barrier [6]. Demonstration of quantum phase-slip (QPS), for which a significant residual resistance will extend far below $T_c$, is non-trivial. The main challenge comes from the need to fabricate a wire of width just a few nanometers, which is beyond the capability of conventional nanofabrication techniques utilizing electron beam lithography or atomic force microscopy. QPS has been established in several systems which have exploited novel nanoengineering techniques, most notably for MoGe nanowires evaporated onto carbon nanotube templates [7], In-based wires formed on ion-beam patterned glass substrates [8], Al nanowires formed on MBE-defined structures [9], and in Al nanowires formed using conventional EBL techniques and reduced by ion beam sputtering of 1 keV Ar+ ions [10]. Recently, new device architectures utilizing coherent QPS for quantum computation and metrology have been explored theoretically [11]. For these purposes the nanowire must be embedded as a functional component within an extended device architecture, incorporating on-chip resistive components. The possibility of engineering nanowires, on the scale of the superconducting coherence length, using conventional fabrication techniques is therefore of importance for the experimental implementation and application of phase-slip phenomena.

This paper reports a study of the low-temperature electronic properties of Nb nanowires fabricated using a focused-ion beam (FIB) to define structures by milling of a thin Nb film with high precision, in a dimension regime where crossover between Josephson and phase-slip behavior is anticipated. As we shall demonstrate, FIB-based fabrication techniques provide access to devices that exhibit phase-slip behavior, aided by Ga-ion poisoning in the periphery of the structure, which reduces the effective electronic dimensions of the superconducting wire significantly below than the physical lithographic limits. The electronic properties observed for a series of nanowires is compared with theoretical models describing thermal and quantum phase-slip processes.

## 2. Experimental details

This paper reports a study of the transport properties of a series of Nb nanowires with width, $W$, in the range 70 nm to 200 nm, and length, $L$, in the range 100 nm to $10^4$ nm. The devices are fabricated by ion-beam milling Nb thin films, of 100 nm thickness and of $T_c$~8.2K, prepared by dc magnetron sputtering on an underlying $SiO_2$/Si substrate. FIB milling is carried out using a crossed-beam FIB/Scanning Electron Microscope (SEM) operating with 30 keV Ga-ions, and is utilized to define the nanowire and bonding pads in the Nb film within the same fabrication step. The SEM, which facilitates imaging of the nanowires (Figure 1a), has been used to estimate their physical dimensions, width, $W$, and length, $L$. A more detailed examination of some of the devices has been performed using High Resolution Transmission Electron Microscopy (HR-TEM) to obtain images of the engineered wires in cross-section. HR-TEM data, shown in Figures 1b and 1c, indicate that the width of the upper part of the wire is considerably smaller than the overall wire width, as determined by SEM imaging. Furthermore, the periphery of the wires is contaminated with Ga, introduced by the ion-beam processing; the TEM images show that in the centre of the upper region of the wire a Nb core free of Ga contamination is present (indicated by the dark region in Figure 1c). The size of this undamaged Ga region is plotted in Figure 1d as a function of the overall wire width determined from SEM imaging. DC current-voltage (*I-V*) measurements were performed between room temperature and 4.2 K, using a Star Cryoelectronics system with nVHz$^{-1/2}$ sensitivity. The resistance of the devices was determined from the *I-V* data at zero-bias-current, so is therefore not influenced by resistive behavior close to the switching current [12], for wires which undergo a transition into a resistive state. For several long wires, with length exceeding 200 nm, this transition was observed to be hysteretic, with the number of steps increasing with increased wire length and reduced width; this behavior will be discussed elsewhere [13].

## 3. Experimental results

The measured variation in $R(T)$ of a series of nanowires with length 100 nm, and varying width, is illustrated in Figure 2. At low temperatures a gradual evolution in the electronic properties of the nanowires is observed with decreasing wire width. Several devices with width exceeding 200 nm, indicated by the open symbols in Figure 2, undergo a superconducting transition at a temperature in the range 7 – 7.3 K, corresponding to the onset of superconductivity in the nanowire. At temperatures below the superconducting transition these devices exhibit Josephson-like behaviour with a critical current that scales appropriately with the device width. A transition at higher temperature, 7.7 K, corresponds to the superconducting transition in the Nb contact pads, which are measured in series with the nanowire [14]. In several narrower devices, indicated by the closed symbols in Figure 2, the superconducting transition is suppressed and a residual resistance is observed at low temperature. In one such device, with width 118 nm (labelled A in Figure 3), the observed $R(T)$ data is suggestive of a phase-slip mechanism extending to low temperature; the behaviour of this device is compared to thermal and quantum phase-slip theory in Section 5. A further five narrow devices, with width $W \leq 100$ nm, exhibit complete suppression of superconductivity. The two narrowest devices, of widths 70 nm and 80 nm, exhibit a negative $dR/dT$ extending to 4 K, suggesting insulating behaviour at low temperature. The side effect of Ga milling is the formation of amorphous damage layers on the walls of the milled trenches with significant Ga concentration (poisoning). Ga-ion poisoning may significantly influence the electronic properties of FIB-defined devices, and it is known that the superconducting properties of Nb are suppressed by the implantation of impurity atoms into interstitial sites [15]. The lateral implantation of Ga ions into the nanowires will be considerable and clearly visible on HR-TEM images in silicon and Nb wires with typical amorphous contrast while the core of the Nb wire remains pristine (Fig 1c). Based on SRIM calculations [16] simulating the implantation of 30 keV Ga-ions into an Nb substrate we estimate a lateral ion straggle of about 35 nm, consistent with measurements made by other groups [1]. We would therefore anticipate damage to the Nb crystal structure over length scales significantly exceeding this range from each surface of the wire, as indicated by the HR-TEM analysis, with the existence of an un-contaminated Nb region at the core of the wire. The observed gradual evolution in the electronic properties with nanowire width of these short devices gives confidence that the observed behavior is not a result of granularity in the films [17]. A similar evolution has been observed in devices with a range of length from 100 nm to $10^4$ nm. The temperature dependence of the resistance for all devices has been compared to theoretical models describing thermal and quantum phase-slip processes in one-dimension. For several devices, of length $\geq 500$ nm, $R(T)$ characteristics have been found to be consistent with thermal phase slip behaviour. These are summarised in Figure 3 along with the corresponding data for device A.

## 4. Theoretical background

Theoretical models have been developed by several authors to explain the temperature dependence arising from thermal and quantum phase slip effects arising in the one-dimensional limit. This section briefly reviews key results, and presents a formulism for modeling the $R(T)$ data in the present experiment; this is implemented in Section 5. Following the approach of Lau [8], the resistance below $T_c$ of each nanowire is fitted to an expression of the form

$$R = \left(R_N^{-1} + (R_{QPS} + R_{LAMH})^{-1}\right)^{-1}, \tag{1}$$

where $R_{LAMH}$ and $R_{QPS}$ are contributions to the resistance arising from thermal and quantum phase-slip respectively, and $R_N$ is the resistance caused by electronic quasi-particles, which represent a parallel conduction channel and can be taken equal to the normal state resistance [4].

Langer, Ambegaokar, McCumber and Halperin (LAMH) [6], derived a model that accounts for phase slip arising due to thermal activation of the order parameter close to $T_c$. They proposed the following formula for $R(T)$ below $T_c$ [6], $R_{LAMH}(T) = (h/4e^2)(\hbar\Omega/k_B T)e^{-\Delta F/K_B T}$, where $\Omega = \sqrt{3\Delta F/4\pi k_B T}\, L/(\xi\tau_{GL})$ is the attempt frequency for activation over the energy barrier, $\tau_{GL} = \dfrac{\pi\hbar}{8k_B(T_c - T)}$ is the GL relaxation time, $L$ is the length of the wire, $\Delta F$ is energy barrier, and $\xi = \xi_0(1-T/T_c)^{-1/2}$ is the GL coherence length. The energy barrier takes the form $\Delta F = 0.83(LR_q/\xi_0 R_n)k_B T_c(1-T/T_c)^{3/2}$, where $R_q = h/(4e^2)$ and $R_n$ is the normal state resistance [7], so that

$$R_{LAMH}(T) = A\left[(1-T/T_C)^{3/2} T_c/T\right]^{3/2} \exp\left(-B(1-T/T_c)^{3/2} T_c/T\right) \tag{2}$$

where $A = R_q \dfrac{8\sqrt{3}}{2\pi^{5/2}} \dfrac{L}{\xi_0}\left(0.83\dfrac{L}{\xi_0}\dfrac{R_q}{R_n}\right)^{1/2}$ and $B = 0.83 LR_q/(\xi_0 R_n)$.

The form of $R_{QPS}(T)$ has been considered theoretically by several authors, who identify a similar form for an exponential dependence on temperature, but with significant variation in the value of the pre-exponential factors [18-20]. Following Chang [20], we write a contribution to the nanowire resistance of the form $R_{QPS}(T) = (\Phi_0/I_0)Ce^{-S/\hbar}$, where $C = \sqrt{S/\hbar} \, L/\left(\pi^{3/2}\tau_{GL}\xi\right)$, $S = H_c^2 \sigma \tau_{GL}\xi/81$, $\Phi_0 = h/(2e)$ and $I_0 = k_B T/\Phi_0$, giving essentially the same temperature dependence as described by the analysis of Lau [7], and Giordano [8]. Simplifying this expression gives

$$R_{QPS}(T) = \alpha_1 A \left[\left(1 - \frac{T}{T_c}\right)^{7/4} \frac{T_c}{T}\right] \exp\left(-\alpha_2 B \sqrt{1 - \frac{T}{T_c}}\right) \quad (3),$$

where $\alpha_1 \approx 1$ and $\alpha_2 \approx 0.03$.

## 5. Discussion

The solid curves in Figure 3 represent fits to theory. Theoretical fitting of the $R(T)$ data of device A was performed with measured values for $W$ and $R_n$, and utilizing the other variables in Table I as fitting parameters. We note that in this case the fitted curves yield unrealistic values for the variable parameter $\xi_0$, of the order of $10^5$ nm as illustrated for example in the first column of Table I, which significantly exceeds the expected coherence length for Nb. For the case of Device A, fitted values given for $\alpha_1$ and $\alpha_2$ also differ significantly from the expected values. An excessive value of the coherence length has been deduced from related studies of thermal phase slip in YBa$_2$Cu$_3$O$_{7-d}$ nanowires by Mikheenko [21], suggesting that the apparent large coherence length can be accounted for by noting that the nanowire consists of an effective superconducting filament that is considerably smaller than the physical dimensions of the FIB-patterned device (as it is also shown in Fig. 1c). Following this approach, the theoretical curves in Figure 3 have been produced using the nanowire width, $W$, as a fitting parameter and setting the coherence length $\xi_0$ to a fixed value. NbGa regions in the periphery of the nanowire, which are expected to have a lower superconducting transition temperature than Nb, may be proximitized by the Nb nanowire, giving rise to differences of $\xi_0$ in both the NbGa region and the Nb region. However, fluctuations would drive the Nb core into a normal state, so that the NbGa periphery would no longer be superconducting by proximity. $\xi_0$ is therefore estimated for the Nb core, with a value of order 10 nm [12]. The results of these fits are summarized in Table I.

Curve AI represents a fit of the form given by (1), incorporating contributions to the nanowire resistance from both thermal (2) and quantum phase-slip (3) processes, in addition to the quasiparticle resistance channel. Curve AII, and curves B-D, consider thermal activation of phase-slip and ignore any contribution to the resistance from quantum phase slip. Curves B-D show the corresponding experimental data for several nanowires with length exceeding 100 nm to be consistent with a residual resistance described by the LAMH model. Evidence of thermal phase-slip has been reported in several superconducting nanowire systems [4] and is well understood; the remainder of this letter therefore focuses on a theoretical description of the device A. We note that attempts to fit the $R(T)$ data for device A at low temperature were not possible if the term $R_{QPS}(T)$ was omitted. The LAMH expression (curve AII) is found to account for our data close to $T_c$, with the fitting parameters given in Table I, but is not consistent with the measured resistance below about 5.4 K. The theoretical model incorporating both thermal and quantum phase-slip processes describes the data over the full temperature range very well, using a value for $T_c$ which is close to that of the Nb film and a nanowire width of 1.8 nm. The evolution of the Nb core width, determined by HR-TEM, with the overall physical width of the wire, determined by SEM (Figure 1d), indicates that wires of width about 100 nm would possess a Nb core size of just a few nanometres. This is consistent with the value for the Nb wire width obtained from theoretical fitting of the electronic properties. The $R(T)$ data suggests that thermal activation of phase-slip is dominant close to $T_c$, as anticipated. However, below a temperature of about 5.4 K, the form of $R(T)$ can no longer be described by a thermal activation model and the experimental data is suggestive of a quantum-activated process at low temperature.

## 6. Summary

The low temperature electronic properties of Nb nanowires, fabricated by FIB milling, have been explored in a dimension regime where crossover between Josephson-like to dissipative, phase-slip behavior is evident as the nanowire width is reduced. Wide nanowires ($W$>200 nm) display Josephson characteristics with a well-defined critical current. Devices with width in the range 100 - 120 nm exhibit behavior consistent with thermal or quantum phase-slip in a central Nb filament of nanoscale dimensions. Nanowires with a physical width less than 100 nm exhibit insulating behavior extending to high temperature, indicating that Ga-ion poisoning leads to complete suppression of superconductivity in this case. The current study demonstrates the unique possibility that FIB processing affords for the fabrication of phase-slip components with nanoscale dimensions embedded in a wire with physical dimensions of order 100 nm. This work was supported by the Australian Research Council. G.C. Tettamanzi acknowledge the Dutch Foundation for Fundamental Research on Matter (FOM) for financial support. C.H.M. acknowledges funding from the EPSRC and the European Commission via project NMP2-CT-2003-505587 "SFINx".

|  | AI (LAMH + QPS) | AI (LAMH + QPS) | AII (LAMH) | B (LAMH) | C (LAMH) | D (LAMH) |
| --- | --- | --- | --- | --- | --- | --- |
| $\xi_0$ (nm) | $2.5 \times 10^5 \pm 1.0 \times 10^5$ | *10* | *10* | *10* | *10* | *10* |
| $W$ (nm) | *118* | 1.8 ± 0.1 | 1.47 ± 0.01 | 0.7 ± 0.1 | 1.0 ± 0.1 | 0.9 ± 0.1 |
| $R_N$ (Ω) | 15 ± 1 | 13 ± 1 | 9.9 ± 0.1 | 2.9 ± 0.1 | 2.5 ± 0.1 | 2.0 ± 0.1 |
| $T_c$ (K) | 7 ± 1 | 8 ± 1 | 9 ± 0.1 | 14 ± 1 | 11.1 ± 0.2 | 12 ± 1 |
| $\alpha_1$ | 189 ± 10 | 0.8 ± 0.1 | - | - | - | - |
| $\alpha_2$ | 55 ± 1 | 0.34 ± 0.01 | - | - | - | - |

**Table I**. Theoretical fitting results illustrated in Figure 3. Parameters indicated in italic text were regarded as being fixed in the corresponding fit.

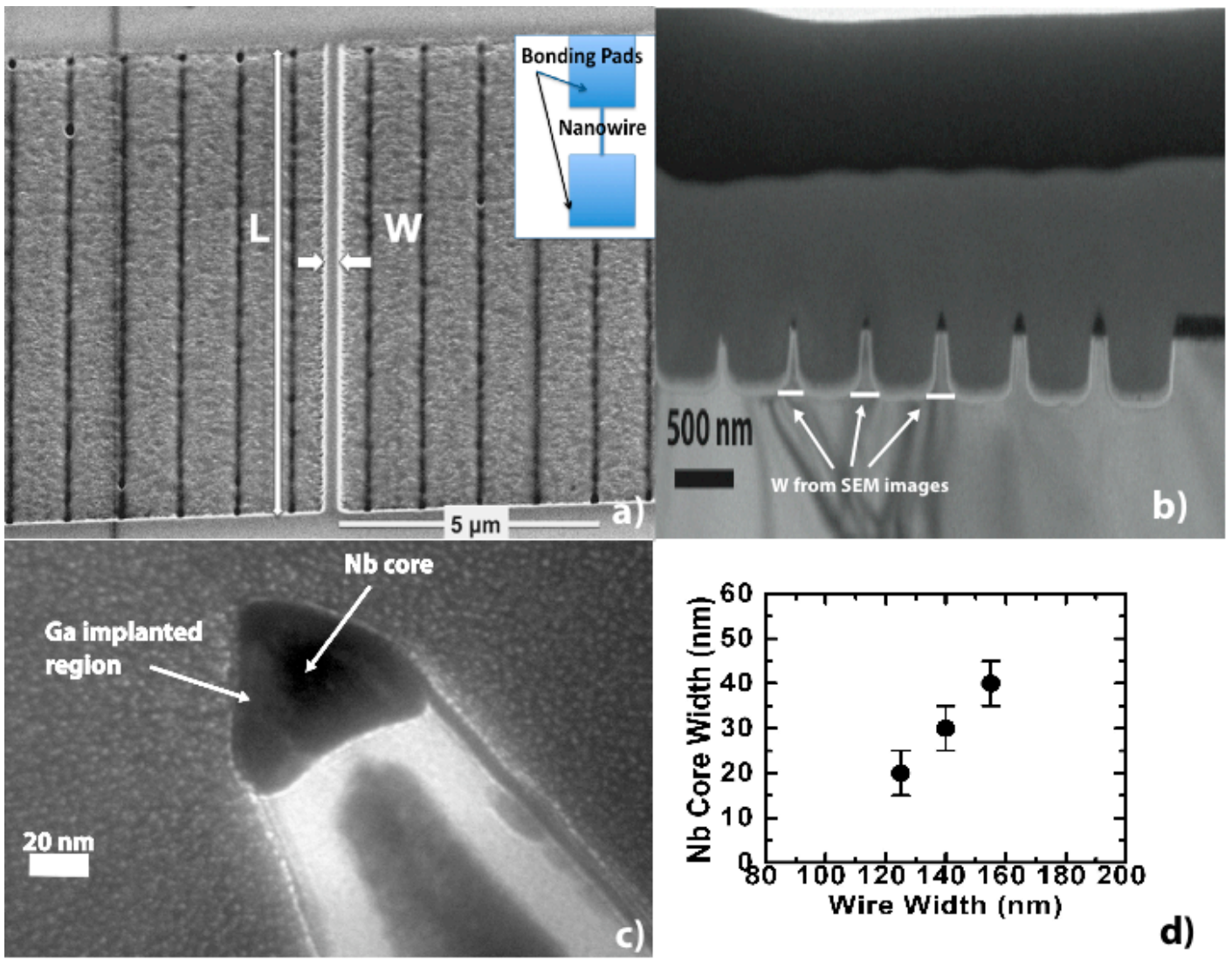

**Figure 1:** a) Scanning electron microscopy image of a long Nb wire. The vertical trenches parallel to the wire, which are in the ion-beam scan direction, arise presumably due to variation in the ion current during processing. In the inset, a schematic of the general structure of the devices is given. b) HR-TEM cross-section of a series of wires fabricated in the similar fashion to the measured wires. The width for these wires when measured by SEM is in the range 100 nm to 200 nm (left to right). c) Enlarged HR-TEM image for a wire of width 130 nm, determined by SEM. A Nb core (in black) of diameter approximately 20 nm, surrounded by Ga-implanted regions is clearly visible. d) Comparison between the wire and Nb core dimensions, determined from SEM and from HR-TEM images. From this graph, a Nb core of a few nanometres in diameter can be estimated for wires that shows a SEM width of ~ 100 nm.

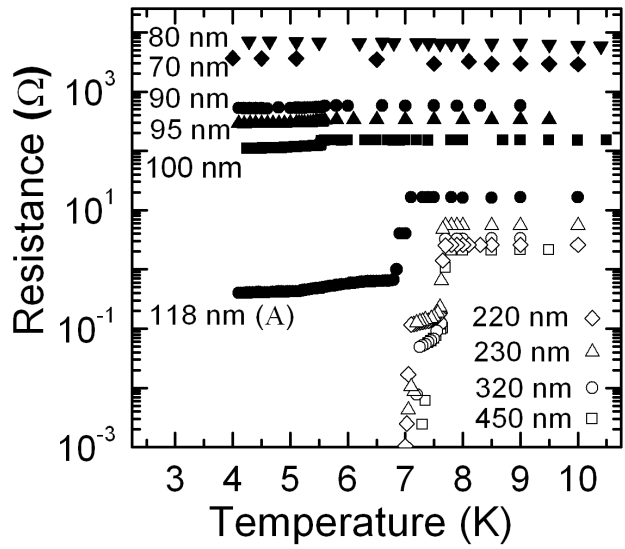

**Figure 2.** $R(T)$ data for several devices with physical length 100 nm. The physical width, determined from SEM images, of each device is indicated. Suppression of superconductivity is evident in the devices represented by closed symbols.

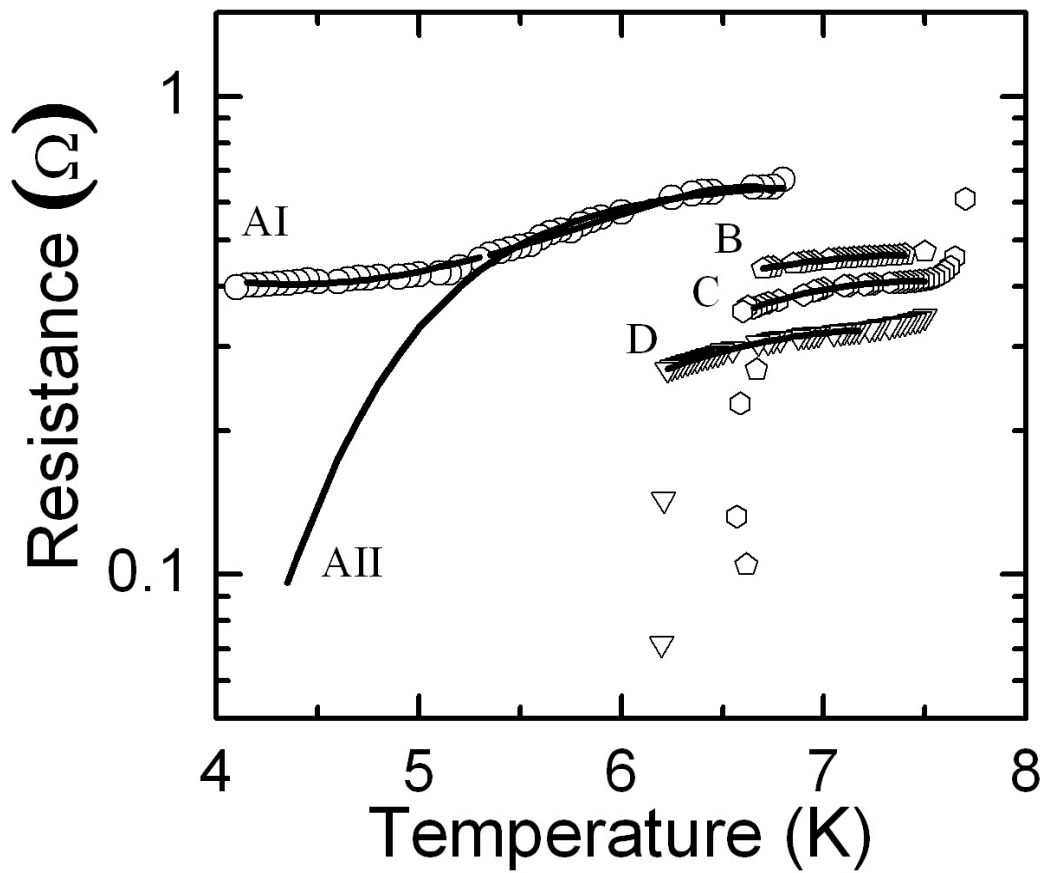

**Figure 3.** Theoretical fits of $R(T)$ for a series of devices exhibiting phase-slip behaviour. AI represents the fit done using both quantum and thermal phase slip theories. AII, B, C and D represent the fits done using only thermal phase slip theory. The symbols represent the experimental data (Device label, Length (nm), Width (nm); A (open circles): 100, 118; B (open pentagons): 5000, 170; C (open hexagons): 10000, 200; D (open triangles): 500, 150). These data are showing trend similar to the one observed by other groups [14]